\documentclass[twocolumn,trackchanges,twocolappendix]{aastex701}

\usepackage{multirow} 
\usepackage{threeparttable}
\usepackage{amsmath} 

\begin{document}

\title{Resolving the Multiple Component Outflows in PG 1211+143:\\
I. The Fe-K Absorption Structure and UFO Forest}

\author[orcid=0000-0003-2161-0361]{Misaki Mizumoto}
\affiliation{Science Education Research Unit, University of Teacher Education Fukuoka, Munakata, Fukuoka 811-4192, Japan}
\email[show]{mizumoto-m@fukuoka-edu.ac.jp}  

\author[orcid=0000-0002-0982-0561]{James N.\ Reeves} 
\affiliation{Institute for Astrophysics and Computational Sciences, Department of Physics, The Catholic University of America, Washington, DC 20064, USA}
\affiliation{INAF, Osservatorio Astronomico di Brera, Via Bianchi 46, I-23807 Merate (LC), Italy}
\email{james.n.reeves456@gmail.com}

\author[orcid=0000-0002-2629-4989
]{Valentina Braito}
\affiliation{INAF, Osservatorio Astronomico di Brera, Via Brera 20, I-20121 Milano, Italy}
\affiliation{Dipartimento di Fisica, Università di Trento, Via Sommarive 14, Trento 38123, Italy}
\affiliation{Institute for Astrophysics and Computational Sciences, Department of Physics, The Catholic University of America, Washington, DC 20064, USA}
\email{valentina.braito@gmail.com}

\author[0000-0001-9735-4873]{Ehud~Behar}
\affiliation{Department of Physics, Technion, Technion City, Haifa 3200003, Israel}
\affiliation{Kavli Institute for Astrophysics and Space Research, Massachusetts Institute of Technology, Cambridge, MA 02139, USA}
\email{behar@physics.technion.ac.il}

\author[orcid=0000-0000-0000-0001]{Chris Done}
\affiliation{Centre for Extragalactic Astronomy, Department of Physics, Durham University, South Road, Durham, DH1 3LE, UK
}
\affiliation{Kavli Institute for the Physics and Mathematics of the Universe (WPI), University of Tokyo, Kashiwa, Chiba 277-8583, Japan
}
\email{chris.done@durham.ac.uk}

\author[orcid=0000-0000-0000-0001]{Kouichi Hagino}
\affiliation{Department of Physics, University of Tokyo, Hongo, Bunkyo-ku, Tokyo 113-0033, Japan
}
\email{kouichi.hagino@phys.s.u-tokyo.ac.jp}

\author[orcid=0000-0000-0000-0001]{Steven B.\ Kraemer}
\affiliation{Institute for Astrophysics and Computational Sciences, Department of Physics, The Catholic University of America, Washington, DC 20064, USA}
\email{kraemer@cua.edu}

\author[orcid=0000-0000-0000-0001]{Gabriele~A.~Matzeu}
\affiliation{Quasar Science Resources SL for ESA, European Space Astronomy Centre (ESAC), Science Operations Department, 28692, Villanueva de la Cañada, Madrid, Spain
}
\email{Gabriele.Matzeu@ext.esa.int}

\author[orcid=0000-0000-0000-0001]{Hirofumi Noda}
\affiliation{Astronomical Institute, Tohoku University, Aramaki, Aoba-ku, Sendai, Miyagi 980-8578, Japan
}
\email{hirofumi.noda@astr.tohoku.ac.jp}

\author[orcid=0000-0000-0000-0001]{Mariko Nomura}
\affiliation{Faculty of Science and Technology, Graduate School of Science and Technology, Hirosaki University, Hirosaki, Aomori 036-8561, Japan
}
\email{nomura@hirosaki-u.ac.jp}

\author[orcid=0000-0000-0000-0001]{Shoji Ogawa}
\affiliation{Institute of Space and Astronautical Science (ISAS), Japan Aerospace Exploration Agency (JAXA), Sagamihara, Kanagawa 252-5210, Japan
}
\email{ogawa.shohji@jaxa.jp}

\author[orcid=0000-0000-0000-0001]{Ken~Ohsuga}
\affiliation{Center for Computational Science, University of Tsukuba, Tennodai, Tsukuba, Ibaraki 305-8577, Japan
}
\email{ohsuga@ccs.tsukuba.ac.jp}

\author[orcid=0000-0000-0000-0001]{Atsushi Tanimoto}
\affiliation{Graduate School of Science and Engineering, Kagoshima University, Kagoshima 890-0065, Japan
}
\email{atsushi.tanimoto@sci.kagoshima-u.ac.jp}

\author[orcid=0000-0000-0000-0001]{Tracey~J.~Turner}
\affiliation{Eureka Scientific, Inc., 2452 Delmer Street Suite 100, Oakland, CA 94602-3017, USA
}
\email{turnertjane@gmail.com}

\author[orcid=0000-0000-0000-0001]{Yoshihiro Ueda}
\affiliation{Department of Astronomy, Kyoto University, Kitashirakawa-Oiwake-cho, Sakyo-ku, Kyoto 606-8502, Japan
}
\email{ueda@kusastro.kyoto-u.ac.jp}

\author[orcid=0000-0000-0000-0001]{Satoshi Yamada}
\affiliation{The Frontier Research Institute for Interdisciplinary Sciences, Tohoku University, Aramaki, Aoba-ku, Sendai, Miyagi 980-8578, Japan
}
\affiliation{Astronomical Institute, Tohoku University, Aramaki, Aoba-ku, Sendai, Miyagi 980-8578, Japan
}
\email{satoshi.yamada@terra.astr.tohoku.ac.jp}

\author{Sreeparna Ganguly}
\affiliation{Dipartimento di Fisica, Università di Trento, Via Sommarive 14, Trento 38123, Italy}
\email{sreeparna.ganguly@unitn.it}

\author{Paolo Somenzi}
\affiliation{Dipartimento di Fisica, Università di Trento, Via Sommarive 14, Trento 38123, Italy}
\email{paolo.somenzi@studenti.unitn.it}

\begin{abstract}
   {We present the initial high-resolution X-ray spectroscopic observations of the Fe-K absorption structure in the luminous nearby quasar PG 1211+143, utilizing the X-ray Imaging and Spectroscopy Mission (XRISM). The primary objective is to characterize the Fe-K absorption features due to Ultra-Fast Outflow (UFO) in this Eddington-luminosity source.} 
   {Observations were conducted with XRISM's Resolve and Xtend instruments, complemented by simultaneous data from XMM-Newton and NuSTAR. A historically bright phase was captured.}
   {The Resolve spectra clearly reveal a prominent P Cygni profile and resolves the Fe-K absorption into six distinct velocity components, ranging from $v = -0.074c$ to $-0.405c$. A similar superposition of multiple UFOs has been reported in PDS~456, suggesting that such a ``UFO forest'' structure may be a common feature of near Eddington-luminosity sources. Some UFO components exhibit narrow line widths of approximately $\sigma \sim 200\,\mathrm{km\,s^{-1}}$, which may indicate that the outflows have reached their terminal velocities, thereby resulting in a smaller velocity shear.
   The mass outflow rate is estimated to be $\dot{M}_\mathrm{out} \sim 1~M_{\odot}~\text{yr}^{-1}$, which is of the order of the Eddington accretion rate. This suggests a physically plausible scenario where the outflow is a significant channel for mass ejection.
   }

\end{abstract}


\section{Introduction}
Active Galactic Nuclei (AGN) are known to profoundly influence the evolution of their host galaxies, particularly through powerful outflows (e.g., \citealt{Harrison2018} and references therein). 
X-ray observations by missions such as XMM-Newton, Suzaku, and NuSTAR have unveiled the existence of Ultra-Fast Outflows (UFOs) (e.g., \citealt{Pounds2003,Reeves2009,Tombesi2010b,gofford2013,matzeu2023,yamada2024}). 
UFOs are distinguished by highly ionized absorption features, such as H-like and He-like iron K-shell transitions, with velocities reaching $|v|>0.03c$. 
The kinetic power and momentum carried by UFOs are often estimated to be comparable to the AGN luminosity or accretion rate, suggesting a crucial role in driving large-scale galactic outflows and establishing the observed $M-\sigma$ relation between Super Massive Black Hole (SMBH) mass and host galaxy velocity dispersion (e.g., \citealt{tombesi2012,king2016,costa2020}).

The X-ray Imaging and Spectroscopy Mission (XRISM; \citealt{XRISM}) carries the Resolve X-ray calorimeter. Resolve boasts an unprecedented energy resolution of $\Delta E\simeq 5$~eV (FWHM) in the 2--12 keV energy band, specifically designed to resolve subtle spectral features in relativistic winds. 
XRISM's initial flagship observation of the luminous quasar PDS 456 (with a BH mass of $\sim5\times10^8\,M_\odot$, accreting at or above the Eddington luminosity) yielded a groundbreaking discovery \citep{XRISM_PDS456}. The Fe K absorption line, previously observed as a single broad feature, was resolved into five distinct discrete velocity components outflowing with a range of velocities from $0.22-0.33c$. Superposition of narrow absorption lines, which is referred to as ``UFO forest'' in this paper, created a single absorption line in the limited energy resolution of CCD detectors.
This observation indicated that the wind structure of PDS 456 is highly inhomogeneous and clumpy, potentially comprising up to a million clumps. The estimated mass outflow rate ($60-300\,M_\odot$~yr$^{-1}$) and kinetic power (exceeding the Eddington luminosity) challenge existing feedback models, suggesting that clumpiness might impede efficient energy transfer to galactic-scale outflows. 
This paradigm shift from a ``single absorber'' to a ``UFO forest'' fundamentally alters our comprehension of wind morphology, directly impacting estimates of kinetic power and feedback models, necessitating a re-evaluation of previous estimations.

PG 1211+143 (redshift $z=0.0809$) is a well-studied luminous quasar with an estimated SMBH mass of $\sim5\times10^7\,M_\odot$ (see Appendix A). 
This object consistently shines at a significant fraction of its Eddington luminosity, making it an ideal target for investigating powerful outflows \citep{Pounds2003,Pounds2016,pounds2006,pounds2025,pounds2007,pounds2009,Reeves2008,Mizumoto2021}. 
Past XMM-Newton and Suzaku observations of PG 1211+143 consistently detected persistent, albeit variable, high-velocity outflows, hinting at the presence of multiple velocity components (e.g., $\sim0.066c$ and $\sim0.13c$ in \citealt{Pounds2016} and  $\sim0.3c$ following an increase of the accretion rate in \citealt{pounds2025}).
The primary motivation for XRISM observations of PG 1211+143 is to leverage Resolve's superior energy resolution to definitively resolve its Fe-K absorption structure. This aims to confirm whether the ``UFO forest'' is a common feature of outflows from Eddington-luminosity AGN and to provide the most detailed spectroscopic characterization of its multi-component structure. This will enable a direct comparison with the PDS 456 results, deepening our understanding of the universality and diversity of powerful AGN winds.

\section{Observations and data reduction}
XRISM observations were conducted from 2024 November 29 to December 4, accumulating a total exposure time of 227~ks (OBSID=201065010). During the observation, the Resolve instrument \citep{kelley2025,ishisaki2025} was operated with the OPEN filter and a closed gate valve. The Xtend instrument \citep{noda2025,uchida2025} was operated in full-window mode.
A significant portion of the XRISM observation was covered by simultaneous observations with other X-ray observatories. NuSTAR \citep{nustar} provided coverage in the hard X-ray band, recording a net exposure time of 83 ks. XMM-Newton \citep{xmm} offered broad-band X-ray and optical/ultraviolet (UV) context, with a net EPIC-pn exposure time of 73 ks. 
Figure \ref{fig:gti} shows detailed information on the overlap between observations.
This paper focuses specifically on the Fe-K energy band, utilizing data above 2.4 keV in the observed frame, which corresponds to 2.59 keV in the rest frame of PG 1211+143. Data from the soft X-ray band, including XMM-Newton/RGS observations, will be presented and analyzed in a forthcoming paper (Reeves et al. in prep.).

Data reduction followed standard procedures, as detailed in the XRISM PDS 456 paper \citep{XRISM_PDS456}. A brief summary of the key steps is provided below.
For Resolve, only High-resolution Primary (Hp) events were used. Data were processed with pre-pipeline version ``005\_003.20Jun2024\_Build8.014'', the pipeline script ``03.00.013.010'', and CALDB version 20250315. Low-resolution Secondary (Ls) events were excluded to prevent contamination from pseudo events \citep{TTWOF}. The Response Matrix File (RMF) was generated using the extra-large ("X") option, incorporating all known instrumental effects. 
Throughout the paper, the model fitting was performed with an energy binning of 5~eV with the C statistics \citep{cash1979}, whereas in some figures the energy binning for the plot has been changed for clarity.
The Non X-ray Background (NXB) for Resolve was generated using the NXB database and modeled accordingly\footnote{\url{https://heasarc.gsfc.nasa.gov/docs/xrism/analysis/nxb/index.html}}. Both source and NXB models were simultaneously fitted to the observed spectrum. In the figures shown in this paper, we present the NXB-subtracted spectra for clarity. See Appendix B for the Resolve NXB spectrum.

For Xtend, the source and background regions were extracted using circular regions with radii of 80 arcsec, positioned on the same CCD chip for accurate background subtraction.
For NuSTAR, the source region was defined as a 50 arcsec radius circle, while the background region was an annulus with inner and outer radii of 150 and 250 arcsec, respectively.
Above 35 keV, the NuSTAR spectrum become background dominated and data were excluded at higher energies.
For XMM-Newton/EPIC-pn, one region with a radius of 35 arcsec was used for the source, and two regions with the same radius for the background.

The count rate is 0.084~cts~s$^{-1}$ (Resolve, 2.4--12~keV), 0.158~cts~s$^{-1}$ (Xtend, 2.4--10~keV), 0.532~cts~s$^{-1}$ (XMM-Newton, 2.4--10~keV), and 0.095~cts~s$^{-1}$ (NuSTAR, 3--35~keV).
The errors correspond to 1$\sigma$ statistical uncertainty throughout this paper.

\section{Results}
\subsection{Overall features}
   
   \begin{figure}[h!]
   \centering
    \includegraphics[width=\columnwidth]{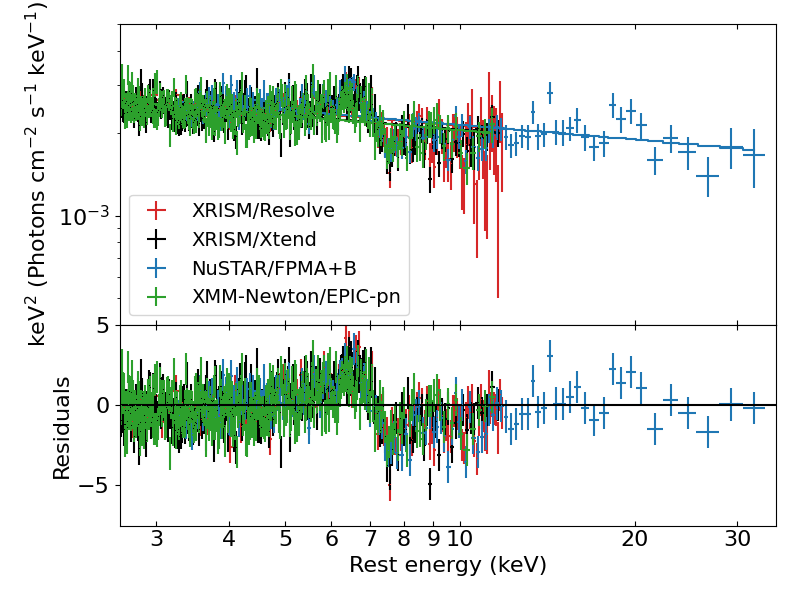}
    \includegraphics[width=\columnwidth]{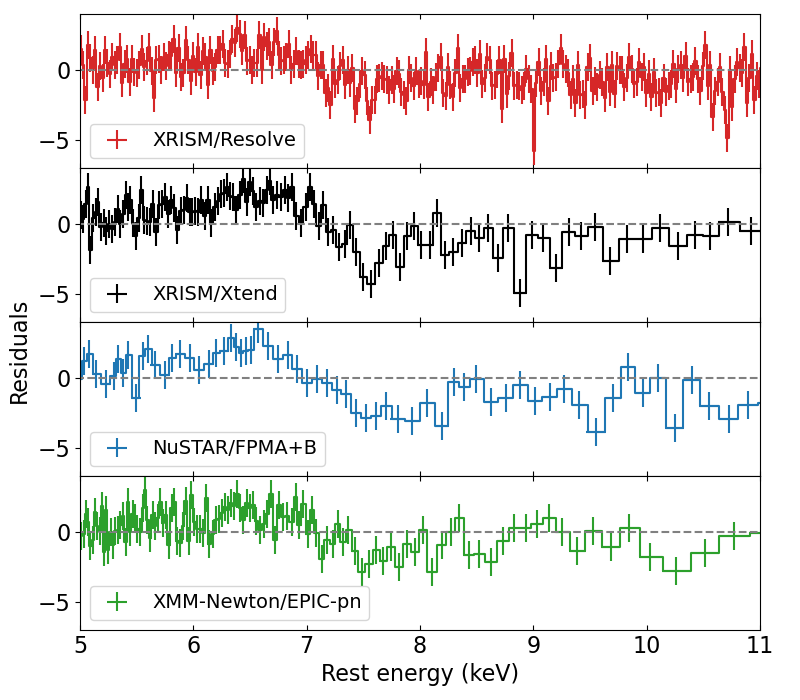}
            \caption{(Upper) Simultaneous XRISM/XMM-Newton/NuSTAR spectra for PG 1211+143. The model line is the power-law continuum without including the Fe-K emission or absorption line features. Error bars correspond to $1\sigma$, and the horizontal axis (energy) is in the rest frame.
            The lower panel shows the residuals against the continuum model.
            (Lower) Inset of the residual plots against a simple powerlaw continuum of photon index of $\Gamma=2.2$. Broad absorption troughs are seen in the Xtend, NuSTAR, and XMM-Newton data between 7--8~keV, and they are resolved as multiple narrow lines in the Resolve data. Other absorption features in the 10--11~keV band are also seen in Resolve, NuSTAR, and XMM-Newton; that it is less prominent in the Xtend data is likely due to poor photon statistics.} A broad emission excess is also seen in all detectors between 5--7 keV.
            
         \label{fig:multiplot}
   \end{figure}

The upper panel of Figure \ref{fig:multiplot} displays the observed full spectra, initially fitted with a single power-law model. Across all detectors (XRISM/Resolve, XRISM/Xtend, XMM-Newton/EPIC-pn, and NuSTAR/FPMA+B), a prominent P Cygni profile and distinct UFO features were consistently observed. The Compton hump, often seen in AGN spectra (see e.g., \citealt{george1991,panagiotou2019}), was not apparent in these observations. 

Crucially, the XRISM/Resolve instrument demonstrated its exceptional energy resolution by resolving the complex UFO absorption lines around 7--8 keV, which remained unresolved by other detectors. This is clearly illustrated in the lower four panels of Figure \ref{fig:multiplot}, which shows the residual plot against a simple powerlaw continuum of photon index of $\Gamma=2.2$. Similar to the results from PDS 456 \citep{XRISM_PDS456}, the absorption features in PG 1211+143 appear to be a superposition of multiple narrow absorption lines with distinct velocities. This unequivocally suggests the presence of a ``UFO forest'' in PG 1211+143.

   
\begin{table}[htbp]
\centering
\caption{Parameter range of the XSTAR table model}
\label{tab:xstar}
\begin{tabular}{lccc}
\hline
Parameter & Range & Grid number & Step \\
\hline
$N_\mathrm{H}$ ($10^{22}$~cm$^{-2}$) & $0.1-200$ & 20  & log\\
$\xi$ (cgs) & $10^3-10^7$ & 20 & log \\
$\sigma_v$ (km s$^{-1}$)& $100-2700$ & 4 & log\\
\hline
\end{tabular}
\end{table}

\begin{figure}[h!]
   \centering
    \includegraphics[width=\columnwidth]{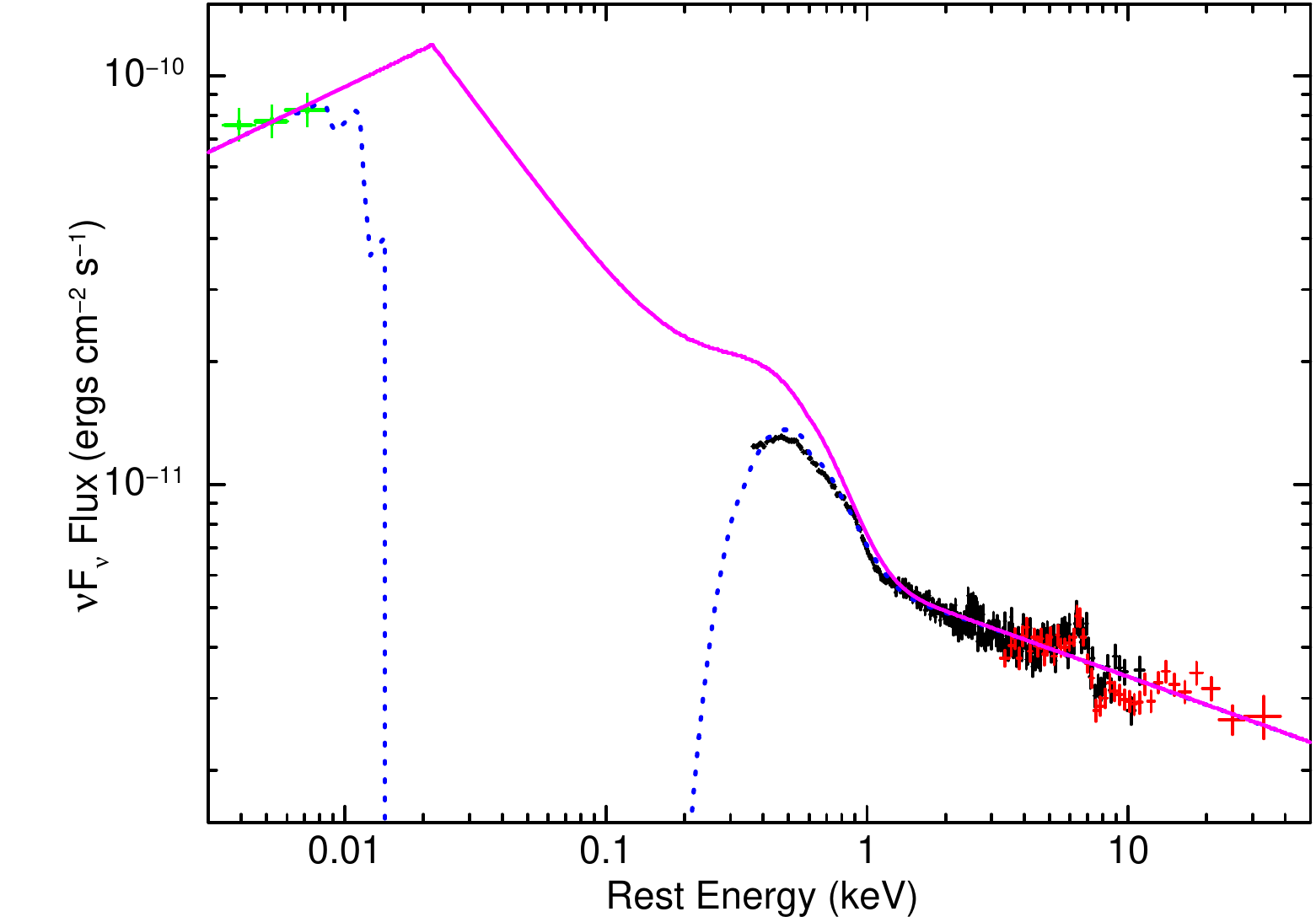}
            \caption{SED in PG 1211+143.
XMM-Newton/EPIC-pn is in black, NuSTAR in red, and XMM-Newton/OM (U, UVW1 and UVW2) photometric points are in green. The OM data are corrected for a Galactic reddening of $E(B-V)=0.035$. 
The solid magenta curve and dotted blue curve are the SED model, but the latter includes the effects of the  neutral Galactic absorption of $N_\mathrm{H}=3\times10^{20}$~cm$^{-2}$.}
         \label{fig:sed}
   \end{figure}

The Spectral Energy Distribution (SED) was determined by XMM-Newton, including the Optical Monitor (OM), and NuSTAR. The SED plot is shown in Figure \ref{fig:sed}, and the best-fit parameter is listed in Table \ref{tab:sed}.
The Galactic absorption of $N_\mathrm{H}=3\times10^{20}$~cm$^{-2}$ or $E(B-V)=0.035$ is assumed. The solid magenta curve is the SED model, parameterized with a double broken powerlaw plus an additional soft excess modeled by the {\tt comptt} model \citep{comptt}. 
The ionizing luminosity from  13.6~eV to 13.6~keV is $L_\mathrm{ion}=(3.1\pm1.0)\times 10^{45}$~erg~s$^{-1}$, and the bolometric luminosity from 1~eV to 200~keV is $L_\mathrm{bol}\simeq6\times 10^{45}$~erg~s$^{-1}$, which is of the order of the Eddington luminosity.
We have captured a historically bright and unabsorbed phase (see also Appendix C), which is brighter compared to the archival XMM-Newton observations from 2001--2014.

\begin{table}[htbp]
\centering
\caption{Best fit parameters for SED}
\label{tab:sed}
\begin{tabular}{llc}
\hline
Component & Parameter & Value \\
\hline
ISM  ({\tt TBabs})& $N_\mathrm{H}$ (cm$^{-2}$) & $3\times10^{20}$ (fix) \\
Reddening ({\tt redden}) & $E(B-V)$ & 0.035 (fix) \\
Continuum ({\tt bkn2pow}) & $\Gamma_1$ & $1.81\pm0.11$ \\
& $E_\mathrm{break,1}$ (keV) & 0.02 (fix) \\
& $\Gamma_2$ & $2.74\pm0.04$ \\
& $E_\mathrm{break,2}$ (keV) & 1.0 (fix) \\
& $\Gamma_3$ & $2.23\pm0.01$ \\
Soft excess ({\tt comptt}) & $kT$ (eV) & $127\pm4$ \\
& $\tau$ & $38^{+6}_{-4}$ \\
\hline
\end{tabular}
\end{table}

\subsection{XSTAR modeling}
To derive the physical parameters of each UFO component, an absorption table model was computed using XSTAR ver 2.59d \citep{xstar}. This model was constructed based on the SED model (the magenta curve in Figure \ref{fig:sed}). The absorption grids were generated to encompass a comprehensive range shown in Table \ref{tab:xstar}, following the methodology outlined by \citet{mochizuki2023}. Solar abundances were assumed for all relevant elements (H, He, C, N, O, F, Ne, Mg, Si, S, Ar, Ca, Fe, Co, Ni) with the xstar default abundance table (xdef)\footnote{\url{https://heasarc.gsfc.nasa.gov/docs/software/xstar/docs/xstarmanual.pdf}}. 

\subsection{Spectral fitting}
The Resolve spectrum, owing to its ability to resolve fine structures, was primarily utilized for the initial determination of model components and their parameters. Subsequently, a simultaneous fit was performed across all available instruments (XRISM/Resolve, XRISM/Xtend, XMM-Newton/EPIC-pn, and NuSTAR/FPMA+B) to assess the statistical significance of each component and obtain robust parameter constraints.

   \begin{figure}[h!]
   \centering
   \includegraphics[width=\columnwidth]{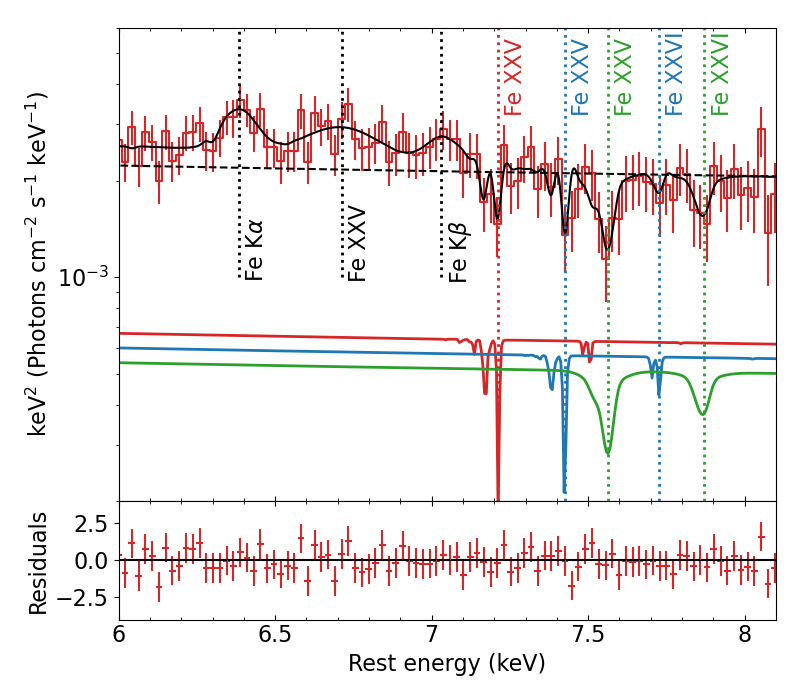}
    \includegraphics[width=\columnwidth]{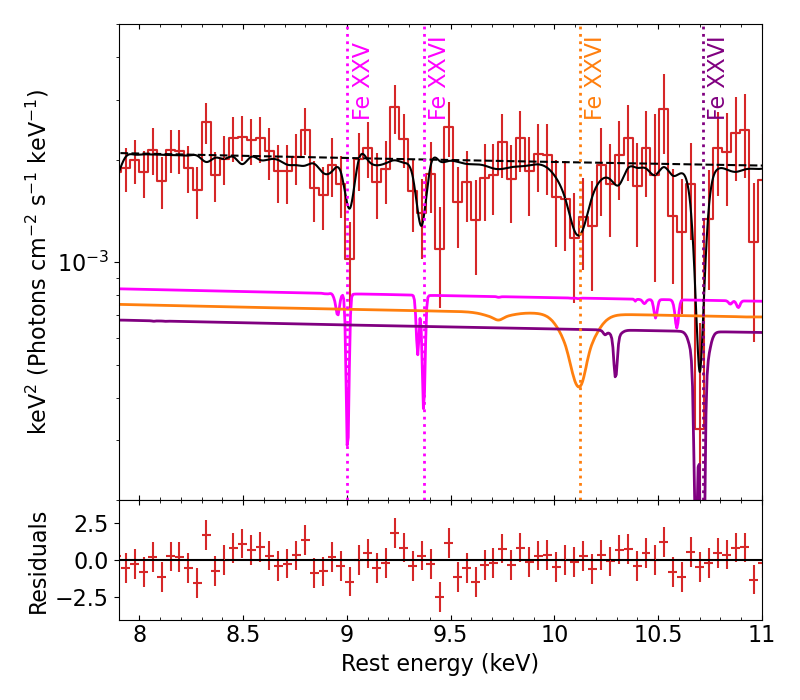}
            \caption{The Resolve spectrum with the best-fit model. The black dotted line shows the emission lines, while the colored, the absorption lines. A spectral model for each absorption component (zones 1--6) is overplotted with some vertical offset; zone 1 = red, zone 2 = blue, zone 3 = green, zone 4 = magenta, zone 5 = orange, and zone 6 = purple.}
         \label{fig:spec1}
   \end{figure}
   
   \begin{figure}[h!]
   \centering
    \includegraphics[width=\columnwidth]{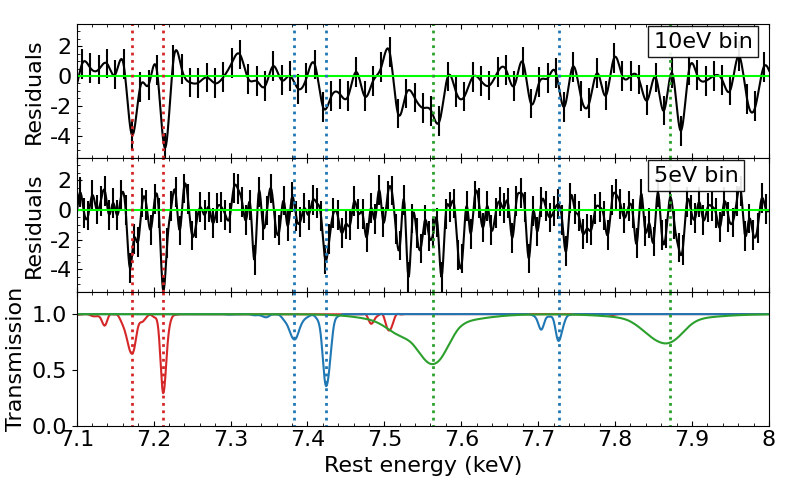}
    \includegraphics[width=\columnwidth]{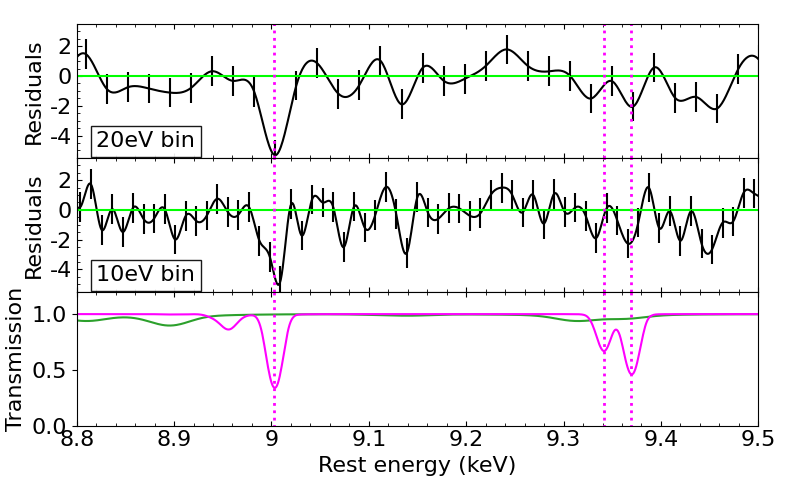}
    \includegraphics[width=\columnwidth]{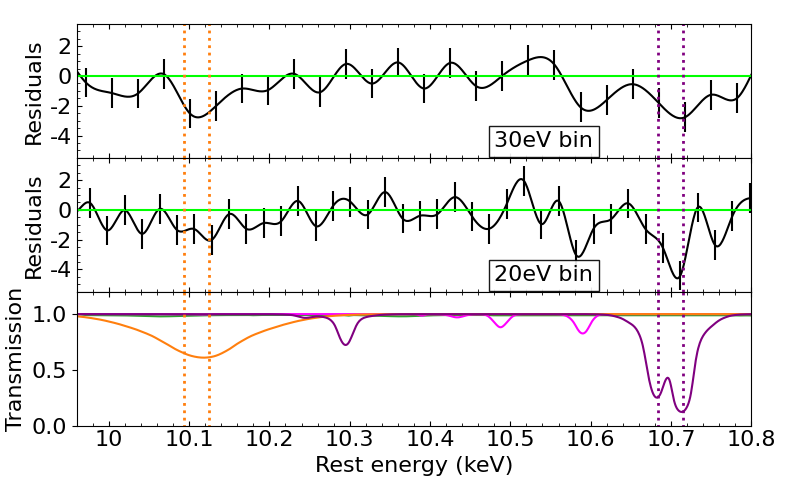}
            \caption{Residual plots against the power law continuum, with multiple energy binning. Each color corresponds to the one in Figure \ref{fig:spec1}. The lower panel shows the transmission of each absorption component. The top three panels show the contribution of zones 1--3, the middle panels zone 4, and the lower panels zones 5--6. The color code is the same as Figure \ref{fig:spec1}.}
         \label{fig:spec2}
   \end{figure}

The comprehensive model included the following:
\begin{itemize}
\item Interstellar medium absorption, fixed according to HI4PI data of $N_\mathrm{H}=3\times10^{20}$~cm$^{-2}$ \citep{HI4PI}. 
\item Three narrow positive Gaussian functions, representing the Fe K$\alpha$, \ion{Fe}{25}, and Fe K$\beta$ emission lines.
\item A single broad positive Gaussian component, accounting for an emission part of the P Cygni profile.
\item Six distinct absorption zones (zone 1 to zone 6), spanning a velocity range from $v/c=-0.07$ to $-0.41$.
\end{itemize}
 The results of the model fitting are shown in Figures \ref{fig:spec1}, \ref{fig:spec2}, and Table \ref{tab:fitting}.

\begin{table*}[htbp]
\centering
\begin{threeparttable}
\caption{Best fit parameters for the simultaneous fitting}
\label{tab:fitting}

\begin{tabular}{llcllc}
\cline{1-3}\cline{5-6}
Component & Parameter & Value & & & Cross norm. \\
\cline{1-3}\cline{5-6}
ISM  ({\tt TBabs})& $N_\mathrm{H}$ (cm$^{-2}$)  & $3\times10^{20}$ (fix) &&Xtend & $0.975\pm0.009$ \\
Powerlaw & $\Gamma$ & $2.215_{-0.011}^{+0.010}$ &&NuSTAR & $1.029_{-0.011}^{+0.006}$ \\
 & norm & $3.30_{-0.06}^{+0.05} \times10^{-3}$  && XMM/EPIC-pn& $0.970_{-0.009}^{+0.008}$\\
\cline{1-3}\cline{5-6}
\multicolumn{2}{c}{Flux (2--10~keV, erg~cm$^{-2}$~s$^{-1}$)} & $6.2\times10^{-12}$ &&&\\
\multicolumn{2}{c}{Luminosity (2--10~keV, erg~s$^{-1}$)} & $1.0\times10^{44}$ &&&\\
\cline{1-3}
\end{tabular}

\vspace{1em}
\begin{tabular}{lccccc}
\hline
 & $E_\mathrm{c}$ (keV) & $\sigma$ (keV)$^{*}$ & norm & $z$ & EW (eV)\\
\hline
Broad (P Cygni)& $5.70_{-0.10}^{+0.11}$ & $0.72_{-0.12}^{+0.11}$ & $1.66_{-0.17}^{+0.16} \times 10^{-5}$ & \multirow{4}{*}{0.0809 (fix)} & $0.24\pm0.02$ \\
Fe K$\alpha$ & $6.379_{-0.010}^{+0.018}$ & $0.06\pm0.02$ & $(3.5\pm0.5) \times 10^{-6}$ & &$44\pm6$\\
\ion{Fe}{25} & $6.71_{-0.02}^{+0.03}$ & $0.12\pm0.04$ & $(4.0_{-0.7}^{+0.5}) \times 10^{-6}$ & &$61^{+8}_{-11}$\\
Fe K$\beta$ & $7.03_{-0.02}^{_+0.04}$ & $0.06$ (tied with Fe K$\alpha$) & $(1.4\pm0.4) \times 10^{-6}$ & &$28\pm8$\\
\hline
\end{tabular}
\vspace{1em}
\begin{tabular}{lccccccc}
\hline
 & $N_\mathrm{H}$ ($10^{22}$~cm$^{-2}$) & $\log\xi$ (cgs) & $\sigma_v$ (km s$^{-1}$)$^{*}$ & $v/c$ & $\Delta C/\Delta\mathrm{dof}$ & $P_\mathrm{null,AIC}$ & $P_\mathrm{null,LEE}$\\
\hline
zone 1 & $1.0\pm0.4$ & 
$4.29_{-0.07}^{+0.12}$ &
$130^{+190}$ & 
$-0.0736\pm0.0003$  &
$-17.62/3$ & $3.0\times10^{-3}$ & 0.03\\
zone 2 & $1.1_{-0.4}^{+0.5} $ &
$4.47_{-0.17}^{+0.22}$ &
$180^{+270}$ &
$-0.102\pm0.003$ &
$-21.44/3$ & $4.4\times10^{-4}$ & 0.005 \\
zone 3 & $3.7\pm1.0$ &
$4.67_{-0.12}^{+0.14}$ &
$1300_{-400}^{+600}$ &
$-0.1205_{-0.0008}^{+0.0012} $ &
$-55.27/3$ & $2.0\times10^{-11}$  & $<0.001$\\
(zone 4)$^{\dagger}$ & $3.0_{-1.1}^{+1.3}$ & 
$4.87_{-0.22}^{+0.16}$ & 
$300_{-90}^{+250}$ & 
$-0.2871_{-0.0003}^{+0.0005}$ & 
$-10.02/3$ & $0.134$ & 0.47\\
(zone 5)$^{\dagger}$ & ${34_{-19}}^\ddagger$ & 
${6.1_{-1.3}}^{*\ddagger}$ & 
2200$^{\S}$ & 
$-0.357_{-0.002}^{+0.003}$ & 
$-5.52/2$ & 0.468 & 0.94\\
zone 6 & ${60_{-30}}^\ddagger $ & 
${6.6_{-0.7}}^{*\ddagger}$ & 
$460\pm280$ & 
$-0.4049_{-0.0003}^{+0.0004}$ & 
$-12.49/2$ & 0.014 & 0.23\\
\hline
\end{tabular}
\begin{tablenotes}[para,flushleft,online,normal] 
\item[$^*$] These values are constrained by the Resolve data and thus fixed in the multi-instrument fitting.\\
\item[$^\dagger$] 
Zones 4 and 5 are marginal detections statistically and thus they are enclosed in parentheses.\\
\item[$^\ddagger$] Only the lower limit can be constrained under the assumption that the absorption line is attributed to \ion{Fe}{26}.\\
\item[\S] Not constrained 
\end{tablenotes}
\end{threeparttable}
\end{table*}

The narrow Fe K$\alpha$ emission line exhibited a width of $\sigma=60\pm20$~eV, which corresponds to a FWHM of $6600\pm2200$~km~s$^{-1}$. This FWHM is approximately three times larger than that of the optical H$\beta$ broad line, FWHM$=1900\pm100$~km~s$^{-1}$ \citep{maha2022}, suggesting that the X-ray Broad Line Region (BLR) can be kinematically distinct and physically closer to the central black hole than the optical BLR.

The width of the Fe K$\beta$ emission line was assumed to be identical to that of Fe K$\alpha$. The observed K$\beta$/K$\alpha$ ratio was remarkably high, reaching 0.4. This unusually large ratio suggests potential contamination of the \ion{Fe}{26} line.  

The initial model fitting to the Resolve spectrum required the inclusion of six absorption components. 
The three slower components have a velocity of $\sim0.07-0.12c$, while the three faster but less significant components have a velocity of $\sim0.28-0.40c$.
They are named zones 1--6, where zone 1 has the slowest velocity and zone 6, fastest.
Figures \ref{fig:spec1} and \ref{fig:spec2} display the color-coded contribution from each of the six absorption zones.

In the subsequent comprehensive multi-instrument fit, the velocity widths of these components were fixed to the values measured from Resolve, and for zones 5 and 6, the ionization parameter ($\log\xi$) was fixed under the assumption of observing H-like iron. 
Zone 1 clearly exhibited \ion{Fe}{25} and \ion{Fe}{24} absorption lines around 7.2 keV in the rest frame. The corresponding \ion{Fe}{26} line was barely detectable, implying a low ionization state.
It should be noted that since these absorption features are observed at an energy just above the Fe K$\beta$ line, an alternative interpretation is lower-ionization K$\beta$ (1s-3p) absorption from iron, albeit at a much lower velocity. In this case, an absorber with $N_\mathrm{H} \sim 3 \times 10^{22} \mathrm{~cm}^{-2}$ and much lower ionization of $\log\xi \sim 2.5$ is required to model either of the 7.2 keV absorption troughs. However, such parameters would produce deep absorption in the soft X-ray band, which is inconsistent with the observed SED (Figure \ref{fig:sed}). It is also inconsistent with the simultaneous RGS spectrum, which at predicts more than 2 orders of magnitude lower absorption at this ionization.
In zone 2, the most prominent line in this component was \ion{Fe}{25}. The \ion{Fe}{24} and \ion{Fe}{26} lines were not strong, and its ionization parameter was marginally higher than that of zone 1.
Zone 3 showed a larger line width ($\sigma_v\sim1300$~km~s$^{-1}$) compared to zones 1 and 2 ($\sigma_v\sim200$~km~s$^{-1}$), and thus produces a broad absorption trough at 7.56~keV due to \ion{Fe}{25} and a weaker trough due to \ion{Fe}{26} at 7.87 keV; see, e.g., the green curves in Figures \ref{fig:spec1} and \ref{fig:spec2}.
Zone 4 showed both \ion{Fe}{25} and \ion{Fe}{26} Ly$\alpha_{1,2}$ lines in the data around 9 keV and 9.4 keV, respectively. Its \ion{Fe}{25} line is particularly prominent in the residual plot (Figure \ref{fig:multiplot}, lower panel). It should be noted that zone 4 exhibits an intermediate ionization parameter and column density (similar to the slower components) while simultaneously possessing a high velocity (comparable to the faster, more highly ionized components).
The zones 5 and 6 are the fastest velocity among all the UFO zones. The absorption lines can be attributed to \ion{Fe}{25} or \ion{Fe}{26}, and are tentatively regarded as the \ion{Fe}{26} line.
Under this assumption, only the lower limit of the ionization parameter and the column density can be constrained. In zone 5, the turbulence velocity cannot be determined within the range of 100--2700~km~s$^{-1}$ sampled in the XSTAR grids (see Table \ref{tab:xstar}).

\begin{figure}[h!]
   \centering
    \includegraphics[width=\columnwidth]{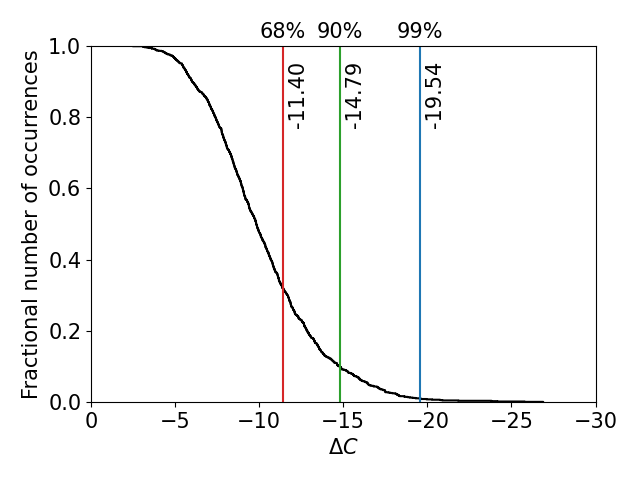}
            \caption{The cumulative histogram for $\Delta C$ as a result of the spectral simulations. Each vertical line show the threshold for each null probability.}
         \label{fig:cumu}
   \end{figure}

The significance, or null hypothesis probability, of each absorption component was evaluated based on the Akaike Information Criterion (AIC; \citealt{akaike74,tan12}) and the Look-Elsewhere Effect (LEE; e.g.,~\citealt{kosec17}; also see \citealt{tombesi2010} for an earlier example of assessing the detection probability of UFO features from Monte Carlo simulations).
To quantify LEE, we simulated 1000 spectra for Resolve, Xtend, XMM-Newton/EPIC-pn, and NuSTAR. These simulations were based on the best-fit model, assuming no UFO absorption. Subsequently, an XSTAR table model was introduced to each set of simulated spectra. The column density and the ionization parameter were treated as free parameters. The line width was held constant at 200~km~s$^{-1}$, a typical value. A redshift scan was performed in the range $z = -0.319$ to $+0.081$ (400 steps) and the corresponding change in the C-statistic was recorded.
The resulting cumulative histogram is presented in Figure \ref{fig:cumu}. Based on this distribution, an absorption component is considered detectable at 1$\sigma$ significance if $\Delta C < -11.40$.
According to both the AIC and LEE criteria, zones 1-–3 and zone 6 were significantly detected. In particular, zones 1--3 are all detected at $>97$\% significance, where zone 3 is the most significant at $>99.9$\%, i.e. no false detections in 1000 simulations.
In contrast, zones 4 and 5 were not statistically significant; consequently, they are enclosed in parentheses in Table \ref{tab:fitting}. It should be noted that both \ion{Fe}{25} and \ion{Fe}{26} are simultaneously detected in zone 4, and while only marginally detected, zone 5 corresponds to a strong depression in the spectrum.

\section{Discussion}

\subsection{Origin of the narrow absorption line width}

\begin{figure}[h!]
   \centering
    \includegraphics[width=\columnwidth]{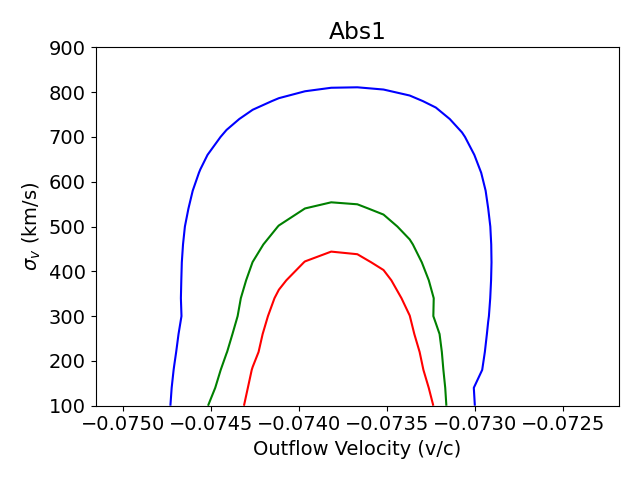}
            \caption{Contour plot for the wind velocity versus the line broadening $\sigma_v$ for zone 1. The red, green, and blue shows 68\%, 90\%, and 99\% confidence level, respectively.}
         \label{fig:contour}
   \end{figure}

Resolve's high-energy resolution enables us to measure the line width with high accuracy. We find that most absorbers are exceptionally narrow, with widths for zones 1, 2, and 4 ($\sim100-300$~km~s$^{-1}$) and even the high-velocity zone 6 ($\sim500$~km~s$^{-1}$) being nearly an order of magnitude smaller than the $\sim1900$~km~s$^{-1}$ reported for PDS 456 (\citealt{XRISM_PDS456}). Only the two broadest absorbers, zones 3 and 5, exhibit a width that reaches a level comparable to that of PDS 456. Figure \ref{fig:contour} shows a contour plot of the wind velocity and $\sigma_v$ for zone 1 in the energy band where the absorption lines appear.

Owing to the calorimeter-resolution spectra provided by Resolve, we can access the line broadening and its mechanism.
Origin of the line broadening can be thermal broadening and shear velocity.
Based on SED (Figure \ref{fig:sed}), the Compton temperature is $T_\mathrm{C}=2.4\times10^6$ K. The thermal broadening of the Fe ion can be calculated as
\begin{equation}
\sigma_{v,\mathrm{thermal}}=\left(\frac{2kT_\mathrm{C}}{m_\mathrm{ion}}\right)^{1/2}=30\,\mathrm{km\,s}^{-1}, \label{eq:thermal}
\end{equation}
where $k$ is the Boltzmann constant, $m_\mathrm{ion}=56m_\mathrm{p}$ is the iron ion mass, and $m_p$ is the proton mass.
The thermal broadening of $30\,\mathrm{km\,s}^{-1}$ is not the main origin of the line broadening.

One of the remaining mechanisms to broaden the absorption line is the shear velocity. 
With the exception of zone 3, the velocity broadening of the Fe absorbers is substantially narrower than in PDS 456 \citep{XRISM_PDS456}. It suggests either that the streamlines we intercept have reached their terminal coasting values \citep{Mizumoto2021}, or that the clumps are small enough such that the velocity shear is relatively small.
It should be noted that even in the wind with $v=-0.4c$ (zone 6) the line width is as narrow as that in zone 1 with $v=-0.07c$. This pattern, which consists of multiple narrow components at different wind velocity, is precisely what physical models of accelerating outflows predict (see Figs.~5, 6 in \citealt{Mizumoto2021}).

To make the gas clumps stable, that is, to keep the gas free from the experience of shocks or fragments, there should be an upper limit to the turbulence velocity, which can be the same order of the sound speed. It can be calculated as 
\begin{equation}
c_s=\left(\frac{2kT_\mathrm{C}}{m_\mathrm{p}}\right)^{1/2}=200\,\mathrm{km\,s}^{-1}. \label{eq:sound}
\end{equation}
This velocity is almost consistent with the observed line broadening except for zone 3, which has substantially greater broadening. Such UFO components can stably exist as gas clumps. 
It could be the case that zone 3 is more extended and thus displays a higher velocity shear, if it is still subject to some acceleration. 

\subsection{Geometry of the wind}
The column density ($N_\mathrm{H}$) and number density ($n$) of the wind gas can be calculated as
\begin{align}
    N_\mathrm{H}&=nd_\mathrm{clump} \\
    n&=L^\prime_\mathrm{ion}/\xi R^2, \label{eq:n}
\end{align}
where $d_\mathrm{clump}$ is the size of each clump, which is assumed to be $10\,R_g$ like PDS 456 \citep{XRISM_PDS456}, $R$ is the radius where the absorption line is made, and $L^\prime_\mathrm{ion}$ is the irradiating luminosity after the correction of Doppler de-boosting.
Based on the observed parameters and Equation (2) from \citet{luminari2020}, the Doppler de-boosting factor $(\Psi$) is calculated, with the assumption of a radial outward motion of the gas (i.e., $\theta=180$~deg in Equation (2) of \citealt{luminari2020}), and $L^\prime_\mathrm{ion}$ is expressed as $\Psi L_\mathrm{ion}$.
As a result, the radius $R$  is
\begin{align*} 
R = & \ 2000 \,R_\mathrm{g} \left(\frac{L^\prime_\mathrm{ion}}{3\times10^{45}\,\mathrm{erg\,s^{-1}}}\right)^{1/2} \left(\frac{d_\mathrm{clump}}{10\,R_\mathrm{g}}\right)^{1/2} \\ &  \left(\frac{M_\mathrm{BH}}{5\times10^7\,M_\odot}\right)^{-1/2} \left(\frac{N_\mathrm{H}}{10^{22}\,\mathrm{cm}^{-2}}\right)^{-1/2} \left(\frac{\xi}{10^5\,\mathrm{cgs}}\right)^{-1/2}. 
\end{align*} 
On the other hand, the wind launching radius ($R_\mathrm{launch}$) can be estimated as
\begin{equation}
    R_\mathrm{launch}/R_\mathrm{g}=2(v_\mathrm{w}/c)^{-2}
\end{equation}
The comparison of these two radii is tabulated in Table \ref{tab:radius}.

\begin{table}[htbp]
\centering
\caption{Wind radius and wind launching radius for each component}
\label{tab:radius}
\begin{tabular}{lcccc}
\hline
& v/c & $\Psi$ & $R/R_\mathrm{g}$  & $R_\mathrm{launch}/R_\mathrm{g}$ \\
\hline
zone 1 & $-0.074$  & 0.74   & 4200 & 370 \\
zone 2 & $-0.102$  & 0.66  & 3000 & 190 \\
zone 3 & $-0.121$  & 0.62  & 1200 & 140 \\
(zone 4) & $-0.287$  & 0.31  & 770 & 24 \\
(zone 5) & $-0.357$  & 0.22  & 50 & 16 \\
zone 6 & $-0.405$  & 0.18 & 20 & 12 \\
\hline
\end{tabular}
\end{table}

The discovery that absorption lines form at radii significantly distant from their initial launching points aligns with the predictions of theoretical models of UV-line-driven disk winds by \citet{Mizumoto2021}. These models predict that winds accelerate as they propagate outwards, leading to the observation of the fastest components at radii larger than those simple escape velocity arguments would suggest. 
It should be stressed that the wind is expected to reach the terminal velocity at $R\sim2000\,R_g$ \citep{Mizumoto2021}. Since $R$ for zones 1 and 2 is larger than $2000\,R_g$, they may already have reached the terminal velocity, explaining the narrow line width discussed in \S4.1. 
In contrast, the broader zone 3 could still undergo some acceleration and would represent an intrinsically faster component than zones 1 or 2.
On the other hand, zones 4 and 6, which also have small $\sigma_v$, may not reach the terminal velocity due to their small radii. Their clump size can be smaller than $d_\mathrm{clump}=10\,R_g$.

   \begin{figure*}[t]
   \centering
    \includegraphics[width=1.7\columnwidth]{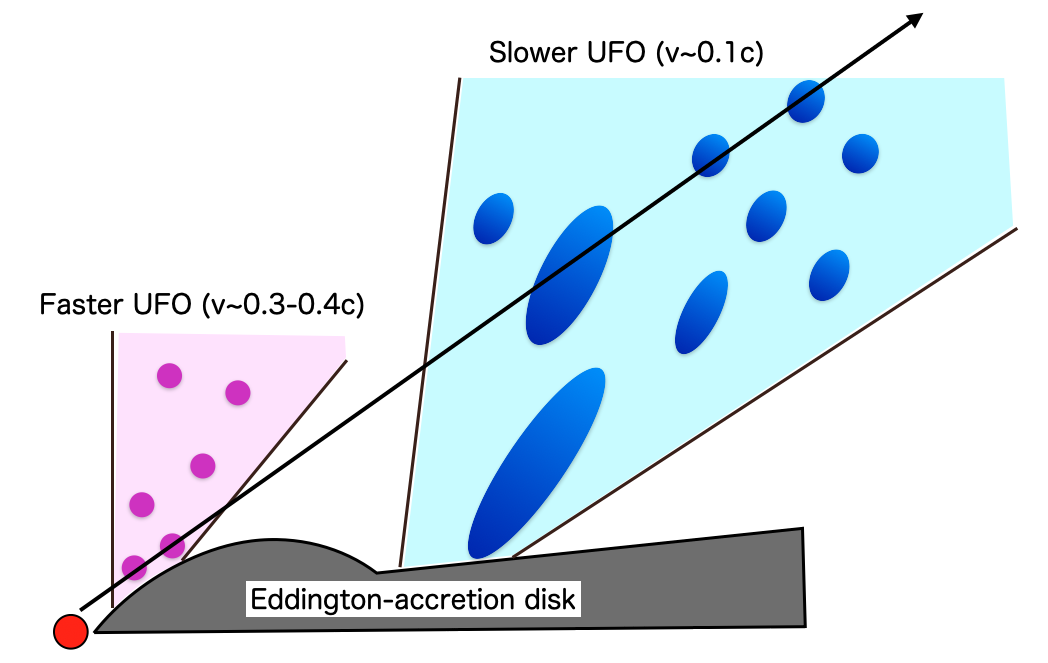}
            \caption{Schematic picture of the UFO in PG 1211+143. The inner components with magenta colors show the faster UFOs, while the outer ones with blue colors show the slower ones. The red circle represents the X-ray corona, and the dark gray area represents the accretion disk. The arrow indicates the line of sight. We expect the wind to be more spatially extended ($f_\mathrm{cov}\sim0.5$, see Section 4.3), but we have narrowed the range for visual clarity.}
         \label{fig:picture}
   \end{figure*}
   
Figure \ref{fig:picture} shows the schematic picture of the UFO in PG 1211+143. The faster inner UFOs may have a small clump size and are located in the inner region. The strong X-ray illumination and magnetohydrodynamics (MHD) process associated within the inner disk could launch the faster winds \citep{fukumura2015}. Within the slower outer UFOs, zone 3 is still accelerated or has an expanded region. Zones 1 and 2 may have reached the terminal velocity. If the Eddington accretion disk is puffed up and shields the strong central X-ray radiation, the slower UFOs can be launched by UV line driving \citep{murray1995,Proga2000,Proga2004,Hagino2015}.
\subsection{Coexistence of Slower and Faster UFOs}

We have found that PG~1211+143 has two distinct velocity groups of UFOs: a slower component (zones 1--3; $v\sim0.1c$) and a faster component (zones 4--6; $v\sim0.3-0.4c$). This contrasts with PDS~456, where only a faster component with $v\sim0.3c$ is observed \citep{XRISM_PDS456}. Here, we investigate the physical origin of this difference.

Slower outflows can be explained by UV line-driven winds, which are launched from the UV-bright region of a standard accretion disk when the effective temperature is optimal, i.e., $T_\mathrm{eff}\sim(0.5-1)\times10^5$~K \citep[see Figure 5 in][]{nomura2021}. In the standard disk model, the effective temperature profile is given by \citep{shakura1973}:
\begin{equation}
T_\mathrm{eff}^4 =
\frac{3GM_\mathrm{BH}\dot{M}_\mathrm{acc}}{8\pi\sigma R^3}
\left[
1-\left(
\frac{R_\mathrm{in}}{R}
\right)^{1/2}
\right], \label{eq:standard}
\end{equation}
where $\dot{M}_\mathrm{acc}$ is the mass accretion rate, $\sigma$ is the Stefan-Boltzmann constant, and $R_\mathrm{in}(\sim6\,R_\mathrm{g})$ is the inner radius of the disk.

For sources accreting near the Eddington limit, such as PG~1211+143 and PDS~456, the photon trapping occurs in the inner region of the disk.
Photons generated within the disk are dragged inward by the accreting gas faster than they can diffuse outward. This traps the radiation, causing energy to be advected into the central object rather than efficiently being radiated away.
This leads to the formation of a ``slim disk.'' The transition occurs at the photon trapping radius, $r_\mathrm{trap}$, described as follows \citep{kato2008}:
\begin{equation}
r_\mathrm{trap}\sim\frac{3}{2}\dot{m}_\mathrm{acc}\left(\frac{H}{R}\right) R_\mathrm{g}, \label{eq:rtrap}
\end{equation}
where $\dot{m}_\mathrm{acc} = \dot{M}_\mathrm{acc}/(L_\mathrm{Edd}/c^2)$ is the Eddington-normalized mass accretion rate and $H/R (\sim1)$ is the scale height of the disk. For a disk to be in the standard regime, its radius must be $r > r_\mathrm{trap}$. Since both PG~1211+143 and PDS~456 have luminosities near the Eddington limit, we can estimate their accretion rates. Assuming a radiative efficiency in the Schwarzschild metric, we have:
\begin{align}
    0.06\dot{M}_\mathrm{acc}c^2&\sim L_\mathrm{Edd} \\
    \dot{m}_\mathrm{acc} &\sim 17.
    \label{eq:dotmacc}
\end{align}

\begin{figure}[h!]
\centering
\includegraphics[width=0.95\columnwidth]{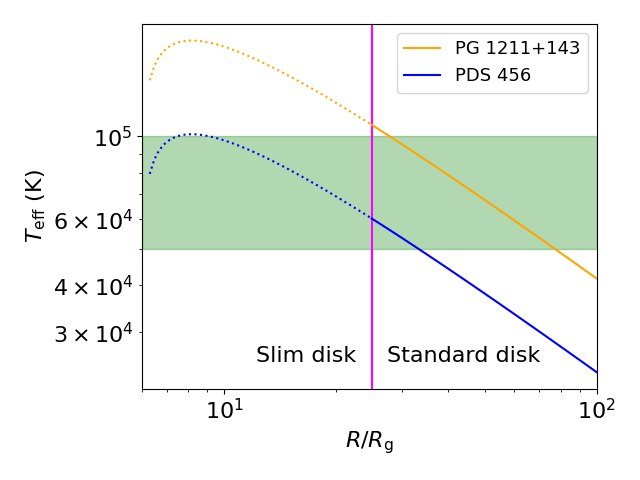}
\caption{The effective temperature ($T_\mathrm{eff}$) of the accretion disk as a function of radius ($R/R_\mathrm{g}$) for PG~1211+143 (orange) and PDS~456 (blue). The green shaded band indicates the optimal temperature range for launching UV line-driven winds ($(0.5-1)\times10^5$~K). The vertical magenta line marks the photon trapping radius ($r_\mathrm{trap}$), which separates the inner slim disk region (left) from the outer standard disk region (right). The curve is shown as a dotted line in the slim disk region, where the temparature profile is not applicable.}
\label{fig:photontrapping}
\end{figure}

Using equations~(\ref{eq:standard})--(\ref{eq:dotmacc}), we can determine the radial extent of the disk region suitable for launching UV line-driven winds. Figure~\ref{fig:photontrapping} illustrates the disk temperature as a function of radius for both AGNs. The key difference is the black hole mass, which is an order of magnitude smaller in PG~1211+143 than in PDS~456. 
A large portion of the outer disk in PG 1211+143 possesses the ideal temperature for launching UV line-driven winds (the slower UFO component), while faster winds can originate from the inner slim disk. In contrast, for PDS 456, although the UV flux itself is high, the launching radius will be smaller.
Contributions from continuum-driven and MHD effects may become significant and a pure UV-driven wind may not be produced in PDS 456.

\subsection{Estimation of mass outflow rate}

The mass outflow rate is described as
\begin{equation}
\dot{M} = 4\pi f_{\text{cov}} f_{\text{vol}} \mu m_p n R^2 v,
\label{eq:mdot_fundamental}
\end{equation}
where $f_{\text{cov}}$ is the covering factor, $f_{\text{vol}}$ is the volume filling factor, $\mu$ is the mean ionic mass in units of the proton mass ($m_p$).
By substituting the term $nR^2$ using the definition of the ionization parameter (Eq.~\ref{eq:n}), the expression can be rewritten as
\begin{equation}
\dot{M} = \mu f_{\text{cov}} \left(\frac{f_{\text{vol}}}{0.01}\right) \left(\frac{v_{\text{out}}}{0.1c}\right) \left(\frac{L^\prime_{\text{ion}}/\xi}{10^{41}\,\text{cgs}} \right)M_{\odot}~\text{yr}^{-1}.
\label{eq:mdot_normalized}
\end{equation}

For zones 1--3 in PG 1211+143, we adopt an ionizing luminosity of $L^\prime_{\text{ion}} = \Psi L_{\text{ion}}  = 2 \times 10^{45}~\text{erg s}^{-1}$ and a typical ionization parameter of $\log \xi = 4.5$. The mean ionic mass is taken to be $\mu \approx 1.4$.
The solid angle covering factor, $f_{\text{cov}}$, is assumed to be 0.5. This value is justified by the observation that the broad emission line strength in PG 1211 is comparable to that in PDS 456 (EW$\sim0.3-0.4$~keV; \citealt{XRISM_PDS456}).

The volume filling factor, $f_{\text{vol}}$, can be estimated based on the multiplicity argument for outflowing clumps, as discussed in the context of PDS 456 \citep{XRISM_PDS456}. It is given by
\begin{equation}
f_{\text{vol}} = 4 M \frac{d_{\text{clump}}}{2R},
\label{eq:fvol}
\end{equation}
where $M$ is the multiplicity (number of clumps along the line of sight) and $d_{\text{clump}}$ is the characteristic size of a clump.
For zones 1--3, we adopt the typical values $R = 2000\,R_g$ and $d_{\text{clump}} = 10 R_g$. With an assumed multiplicity of $M=3$, Eq.~(\ref{eq:fvol}) yields $f_{\text{vol}} = 0.03$.

Substituting the derived and assumed parameters into Eq.~(\ref{eq:mdot_normalized}), we obtain an order-of-magnitude estimate for the mass outflow rate of
\begin{equation}
\dot{M}_\mathrm{out} \sim 1~M_{\odot}~\text{yr}^{-1}.
\end{equation}
With $v=0.1c$, the outflow momentum rate is $\sim2\times10^{35}\,\mathrm{dyne}\sim L_\mathrm{bol}/c$, and the kinetic power is $\sim6\times10^{44}\,\mathrm{erg\,s}^{-1}\sim 0.1\,L_\mathrm{bol}$.

The mass outflow rate is on the order of the Eddington accretion rate for the black hole mass of PG 1211+143 ($M_{\text{BH}} = 5 \times 10^7 M_{\odot}$), suggesting a physically plausible scenario where the outflow is a significant channel for mass ejection.
It is also consistent with previous estimates, e.g., from the analysis of multi-epoch XMM-Newton observations \citep{reeves2018}.

It should be noted that this estimate does not account for the potential higher velocity components of the outflow. However, such zones (with the possible exception of the anomalous zone 4) are typically characterized by a much higher ionization parameter ($\log \xi \approx 6$). Consequently, even with a larger outflow velocity, their contribution may not dramatically increase the total mass outflow rate.

\section{Conclusion}
This paper reports the initial results from high-energy resolution spectroscopic observations of the luminous quasar PG 1211+143 using XRISM/Resolve. We have successfully resolved previously broad Fe-K absorption features into a complex ``UFO forest'' comprising six distinct narrow velocity components. These components span a wide range of outflow velocities, from $v/c=-0.07$ to $-0.41$, and exhibit varying ionization states and column densities. The presence of a prominent P Cygni profile and broad emission lines further characterizes this powerful outflow. Our analysis also confirms that the X-ray Broad Line Region (BLR) is significantly closer to the central black hole than the optical BLR, as evidenced by the broader Fe K$\alpha$ emission line compared to H$\beta$. 

The discovery of the ``UFO forest'' in PG 1211+143, in conjunction with its previous detection in PDS 456, strongly suggests that this intricate multi-component outflow structure is a common characteristic of powerful Eddington-luminosity AGN. This implies underlying universal physical mechanisms responsible for launching and accelerating these extreme winds. 
It should be noted that the UltraLuminous InfraRed Galaxy (ULIRG) with near-Eddington accretion, IRAS 05189--2524, also shows such ``UFO forest'' (Noda et al.\ accepted in ApJL).
The absorption line width in PG 1211+143 is as narrow as $\sim200$~km~s$^{-1}$. This suggests that the wind reaches the terminal velocity with small shear.
Our derived absorption line formation radii, located far from the launching points, align with the theoretical predictions of \citet{Mizumoto2021}, reinforcing the need for physical models to accurately constrain the outflow energetics. 
The mass loss rate is $\dot{M}_\mathrm{out} \sim 1~M_{\odot}~\text{yr}^{-1}$, which is on the order of the Eddington accretion rate.

This paper serves as the first in a series of papers. Future work will involve detailed analysis of soft X-ray band data, particularly from XMM-Newton/RGS observations, to further characterize the multi-phase nature of the outflow across a broader energy range. Continued theoretical efforts are essential to fully explain the observed complex wind structures and P Cygni profiles, the specific kinematic and ionization properties of individual components, and the observed differences among high-Eddington sources.

\begin{acknowledgements}
Part of this work was supported by 
JSPS KAKENHI Grant Number JP21K13958 (MM),
JP21K13963, JP24K00638 (KH),
JP19K21884, JP20H01947, JP20KK0071, JP23K20239, JP24K00672, JP25H00660 (HN),
JP20K14525 (MN),
JP24K17104 (SO),
JP20H01946 (YU),
JP23K13154 (SY),
NASA grants 80NSSC25K7845 and 80NSSC22K0474 (JNR),
STFC through grant ST/T000244/1 (CD),
Yamada Science Foundation (MM), 
Exploratory Research Grant for Young Scientists, Hirosaki University (MN),
the Kagoshima University postdoctoral research program
(KU-DREAM, AT),
and  JSPS Core-to-Core Program (grant number:JPJSCCA20220002).
\end{acknowledgements}

\begin{contribution}
MM was responsible for writing and submitting the manuscript.
MM, JNR, and VB have led the data reduction, analysis and discussion. JNR and MM are the PIs of the XRISM observation.
EB, CD, KH, SBK, GM, HN, MN, SO, KO, AT, TJT, YU, and SY contributed to the discussion and review of the manuscript.


\end{contribution}

\facility{XRISM \citep{XRISM}, XMM-Newton \citep{xmm}, NuSTAR \citep{nustar}}

\software{HEASoft/FTOOLS \citep{heasoft,ftools}
          }

\appendix
\restartappendixnumbering
\section{BH mass}

The black hole mass in PG 1211+143 was estimated using the reverberation mapping, to be $(14.6\pm4.4)\times10^7\,M_\odot$ (large uncertainty due to low variability; \citealt{peterson2004}), $4.05^{+0.96}_{-1.21}\times10^7\,M_\odot$ (using mean spectra; \citealt{kaspi00}), $2.36_{-0.70}^{+0.56}\times10^7\,M_\odot$ (using rms spectra; \citealt{kaspi00}).
In this paper we adopt the mass of $5\times10^7\,M_\odot$.

\section{Resolve NXB spectrum}
\begin{figure}[h!]
   \centering
    \includegraphics[width=0.95\columnwidth]{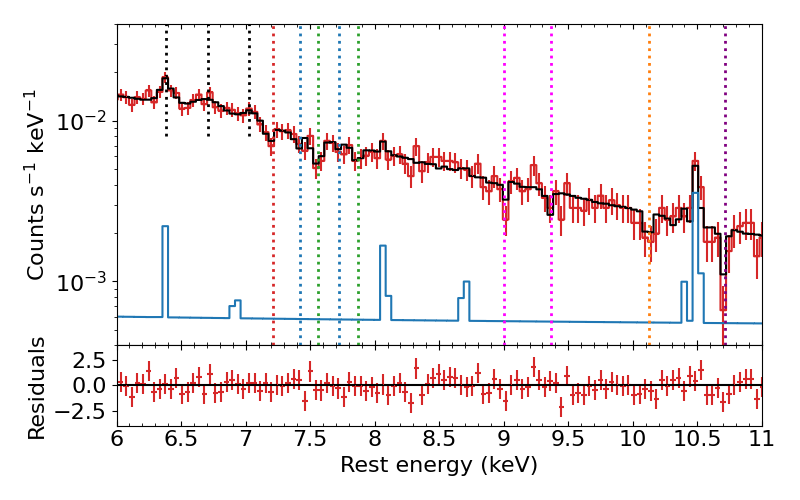}
            \caption{The Resolve spectrum (red) with the NXB model (blue). The black line shows the best-fit model. The color code is the same as Figure \ref{fig:spec1}. The lower panel shows the residuals.}
         \label{fig:nxb}
   \end{figure}

Figure \ref{fig:nxb} shows the NXB model used in this study. The continuum level is about $6\times10^{-4}$~cts~$\mathrm{s}^{-1}$~keV$^{-1}$. This is one tenth of the source continuum level at $\sim9$~keV, and one third at $\sim11$~keV. The Mn K$\alpha$ line at 5.9 keV (6.4 keV at the rest energy of PG 1211+143) overlaps the celestial Fe K$\alpha$ line, but its width and EW are $\sim5$\% and $\sim10$\% of the Fe K$\alpha$ line, respectively. We can fully resolve the NXB Mn K$\alpha$ with the celestial Fe K$\alpha$. The most remarkable NXB feature is the Au lines at 9.6 and 9.7~keV (10.4 and 10.5~keV at the rest energy of PG 1211+143). This is between the absorption features due to zones 5 and 6, and thus there is little effect on the UFO absorption line study.

\section{Time variability}
\begin{figure}[h!]
   \centering
    \includegraphics[width=0.8\columnwidth]{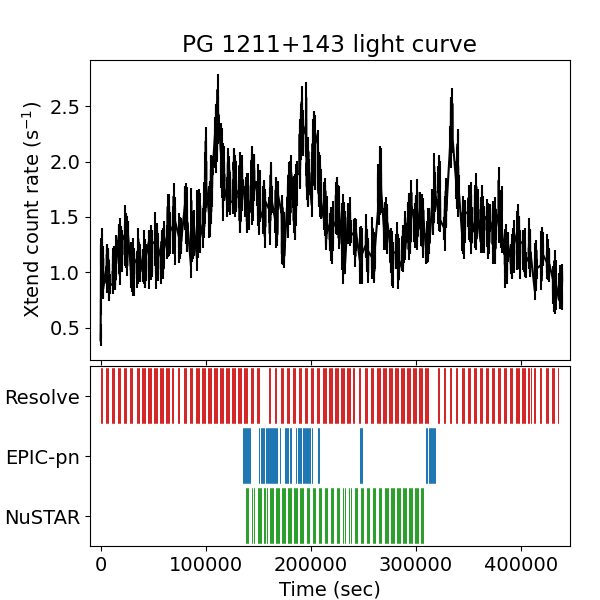}
            \caption{Xtend light curve and GTI in each detector.}
         \label{fig:gti}
   \end{figure}

Figure \ref{fig:gti} shows the short-time variability within our XRISM observation. The Xtend light curve is shown, revealing several flares where the count rate approximately doubled. Although the XMM-Newton and NuSTAR observations were conducted during the XRISM observation period, they did not provide complete coverage. Notably, the XMM-Newton observation appears to have been performed during a time of relatively high count rates. The flux and spectral variations during the observation period will be addressed in our subsequent paper.

\begin{figure}[h!]
   \centering
    \includegraphics[width=0.7\columnwidth,angle=270]{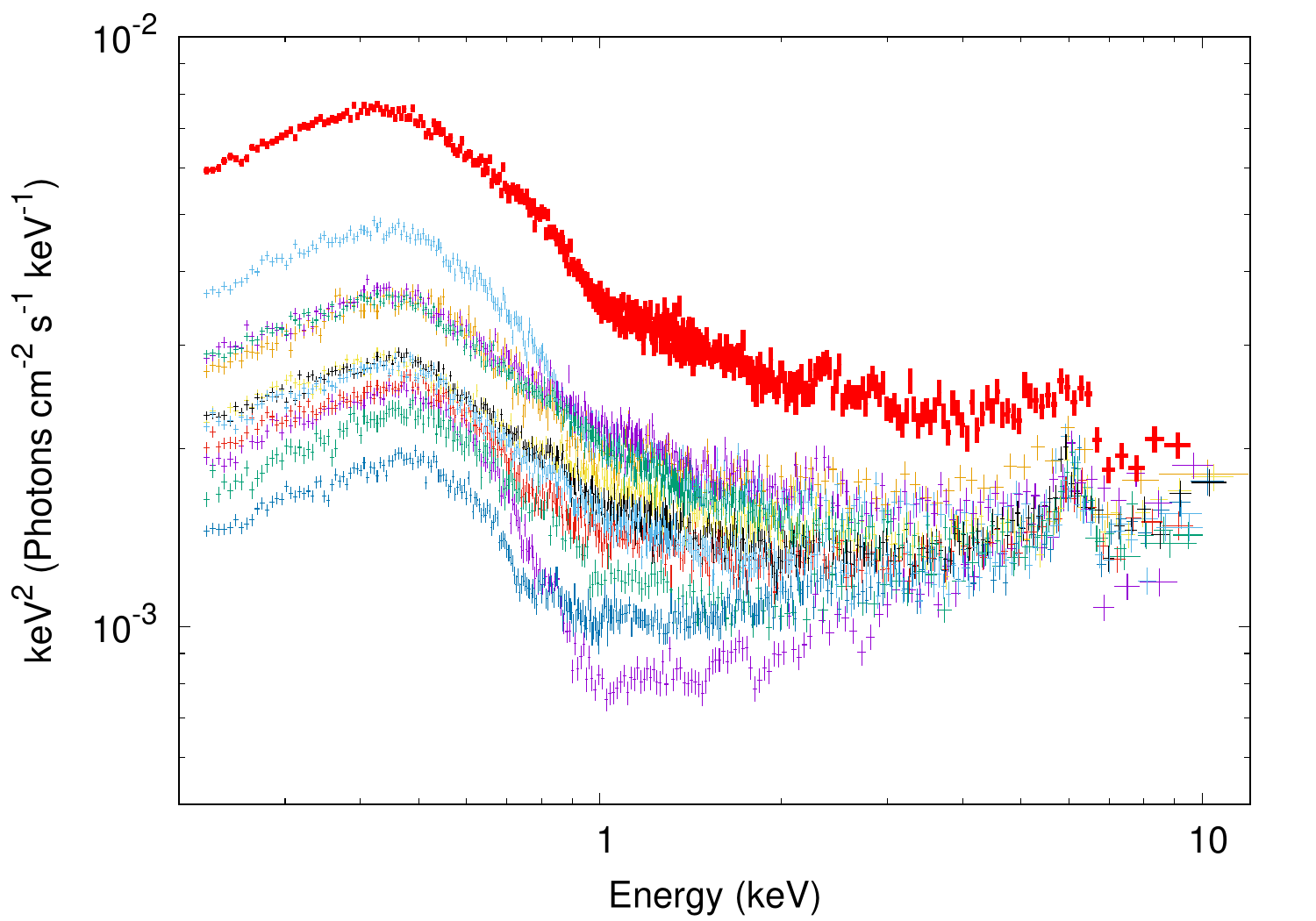}
            \caption{X-ray spectra in PG 1211+143 with all the archival XMM-Newton/EPIC-pn data. Our observations are shown in red.}
         \label{fig:xmm}
   \end{figure}

Figure \ref{fig:xmm} shows the long-term variability using all archival XMM-Newton/EPIC-pn data. Our observation, indicated in red, is the uppermost data set in the figure. We have observed the source during its brightest phase to date, characterized by weaker absorption and the least prominent Fe-K emission line.

\bibliography{main}{}
\bibliographystyle{aasjournalv7}



\end{document}